\documentclass[12pt]{iopart}
\usepackage{graphicx}
\usepackage{epsfig}
\begin{document}

\title[Measurement of the magnetic field inside the holes of a drilled bulk HTS]{Measurement of the magnetic field inside the holes of a drilled bulk high-Tc superconductor}
\author{Gregory P~Lousberg$^{1,4}$, Jean-Fran\c{c}ois~Fagnard$^{1}$, Jacques G Noudem$^{2}$, Marcel~Ausloos$^{3}$, Benoit~Vanderheyden$^{1}$, and Philippe~Vanderbemden$^{1}$}
\address{$^1$ SUPRATECS Research Group, Dept. of Electrical Engineering and Computer Science (B28), University of Li\`ege, Belgium}
\address{$^2$ CRISMAT-ENSICAEN, Caen, France}
\address{$^3$ SUPRATECS (B5a), University of Li\`ege, Belgium}
\address{$^4$ FRS-FNRS fellowship}
\ead{gregory.lousberg@ulg.ac.be}

\begin{abstract}
\linespread{2}
   \selectfont

We use macroscopic holes drilled in a bulk YBCO superconductor to probe its magnetic properties in the volume of the sample. The sample is subjected to an AC magnetic flux with a density ranging from $30~\mathrm{mT}$ to $130~\mathrm{mT}$ and the flux in the superconductor is probed by miniature coils inserted in the holes. In a given hole, three different penetration regimes can be observed: (i) the shielded regime, where no magnetic flux threads the hole; (ii) the gradual penetration regime, where the waveform of the magnetic field has a clipped sine shape whose fundamental component scales with the applied field; and (iii) the flux concentration regime, where the waveform of the magnetic field is nearly a sine wave, with an amplitude exceeding that of the applied field by up to a factor of two. The distribution of the penetration regimes in the holes is compared with that of the magnetic flux density at the top and bottom surfaces of the sample, and is interpreted with the help of optical polarized light micrographs of these surfaces. We show that the measurement of the magnetic field inside the holes can be used as a local characterization of the bulk magnetic properties of the sample.

\end{abstract}

\submitto{\SUST}


\maketitle
\linespread{2}
   \selectfont 
\section{Introduction}

Single domains of bulk melt-processed RBCO (where R denotes a rare earth) have a significant potential for use in permanent-magnet engineering applications~\cite{A1,A2,A3,A4,A5,A6}, due to their ability to trap relatively large magnetic fields~\cite{B1,B2}. Bulk RBCO samples are usually made with either a parallelepipedic or a disk geometry~\cite{C1,C2,C3,C4} and have typical lengths or diameters of $20-50~\mathrm{mm}$ and thicknesses of $5-10~\mathrm{mm}$. With such large volumes, new specifically designed magnetic measurements systems have been developed for characterizing the magnetic properties of bulk RBCO. Magnetic characterization methods applicable to superconducting samples of large size can be divided in two categories : (i) \textit{volume measurements}, i.e. probing a magnetic response that is averaged over the whole sample volume, or (ii) \textit{surface measurements}, i.e. measuring the distribution of the magnetic field at the surface of the sample. The first category of methods includes DC magnetization measurements~\cite{D1,D2} and AC susceptometry~\cite{E1,E2,E3,E4}. The second category involves measuring the distribution of the magnetic field at the surface of the material, either by scanning a miniature Hall sensor over the sample surface~\cite{F1,F2}, or by using an array of (fixed) Hall sensors~\cite{G1}. In these experiments, the samples are either subjected to a uniform applied magnetic field, for flux penetration characterization, or permanently magnetized, for trapped flux measurements. Variants consist in using a small permanent magnet that magnetizes the sample locally during the scanning process (the so-called \textit{magnetoscan} technique~\cite{G2,G3,G4}), or in scanning the surface with a miniature coil system generating an AC magnetic field and detecting the local inductive response~\cite{G5}. The magnetic signal obtained in \textit{surface measurements} can be then used to estimate physical quantities of practical interest, e.g. the critical current density~\cite{C3,C4,D1,D2}, or the levitation forces~\cite{G6}. Since melt-processed materials may also contain numerous defects that impede the current flow (cracks, subgrain boundaries, oxygen deficient regions, $\ldots$), the measurements are also of prime importance to assess the sample homogeneity, to locate cracks or multiple superconducting grains and their deleterious influence on the material performances~\cite{G2,G6,H1}. 

Recently, bulk HTS cylinders with artificial columnar holes have been synthesized to enhance the chemical, thermal, and mechanical properties of HTS~\cite{B1,H2,H3,H4}. These drilled structures offer an alternative method for characterizing the magnetic properties of bulk HTS which is based on a local probing of the magnetic flux with coils inserted \emph{inside} the holes. To the best of our knowledge, such a method has not been elaborated yet, although it can potentially be very effective in assessing both the extrinsic superconducting properties of the sample and the effects of the microstructure of the material. Moreover, as drilled HTS are very attractive for pulsed-field magnetization~\cite{Drilled2}, the measurement of the magnetic flux threading the holes during the pulse duration could bring essential information on the magnetic flux propagation in the volume of the sample during a pulsed-field experiment. 

The purpose of this work is to carry out measurements of the magnetic field threading the holes of a drilled sample subjected to an AC magnetic field, by using micro coils placed inside the holes. The results are correlated with Hall probe scans on the surfaces of the sample. Such a method combines the advantages of \textit{volume} and \textit{surface} measurement methods, since the magnetic flux density is probed locally in the volume of the superconductor. 

The paper is organized as follows. In section~\ref{s:sample}, we describe the characteristics of the YBCO drilled sample used in this work. Section~\ref{s:exp} is devoted to the experimental method used for probing the magnetic field inside the holes of a YBCO drilled cylinder. The description and discussion of the penetration of an AC magnetic field inside the holes are then presented in section~\ref{s:result}.  

\section{YBCO drilled sample}
\label{s:sample}

\begin{figure}[b]
\center
\includegraphics[width=4cm]{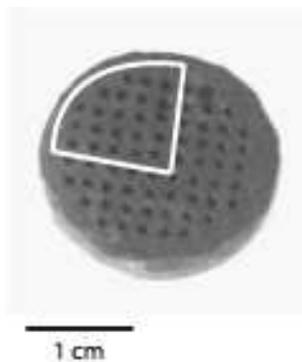}\caption{Photograph of the YBCO drilled sample used in this work. The white line delimits the quarter of the sample under consideration in this paper. }\label{micro}
\end{figure}

Measurements of the magnetic flux were carried out inside the holes of a YBCO drilled cylinder containing 68 artificially patterned holes. The sample was produced at CRISMAT (Caen, France) by the top-seeded melt-textured growth process from a drilled preform~\cite{J2}. The details of the sample preparation can be found in Ref.~\cite{J1}. The sample has a diameter of $20~\mathrm{mm}$ and a height of $8~\mathrm{mm}$. The holes are parallel to the $c$-axis and are arranged on a regular square pattern; the macroscopic holes have a diameter of $0.7~\mathrm{mm}$ and are separated by a distance of $\sim 1.5~\mathrm{mm}$. A picture of the sample is shown in Figure~\ref{micro}. The critical current density of the YBCO material is $J_c\approx1000~\mathrm{A/cm}^2$ at $T = 77~\mathrm{K}$ and $B = 100~\mathrm{mT}$. In order to reduce the number of thermal cycles, we have only analysed the field in the holes belonging to the quarter of the sample delimited by the white lines.

The magnetic properties of the sample have been characterized first with usual Hall probe scans over both surfaces. The spatial distribution of the magnetic field above the sample was measured by scanning a miniature probe, fixed to a motor-driven $xy$ micro-positioning stage, over the sample surface~\cite{H2}. The active area of the Hall probe, sensitive to the component of the local field, which is perpendicular to the surface, is $0.05\times0.05~\mathrm{mm}^2$. The gap between the Hall probe and the sample surface is $0.2~\mathrm{mm}$. The Hall probe was  moved across either the top or bottom surfaces with a step size of $0.5~\mathrm{mm}$ in $x$ and $y$ directions. In view of comparing the Hall probe mapping with measurements of the AC magnetic field inside the holes, we measured the penetration of the DC magnetic field ($100~\mathrm{mT}$) applied after a zero-field cooled process (down to $77~\mathrm{K}$) and kept constant during the Hall probe mapping experiment.

\begin{figure}[t]
\center
\includegraphics[width=10cm]{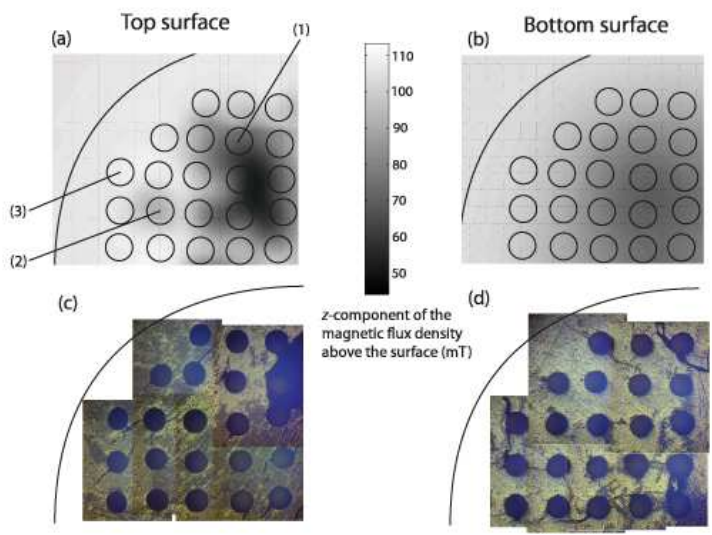}
\caption{(a,b)- DC Magnetic flux density distribution measured in a Hall probe mapping experiment at $0.2~\mathrm{mm}$ above the top (a) and bottom (b) surface of the quarter of the sample under consideration. The sample is subjected to a DC uniform magnetic flux density of $100~\mathrm{mT}$ during the measurement. A grid showing the position of the holes is superimposed. (c,d)- Picture of the top (c) and bottom (d) surface of the quarter of sample acquired with a micrograph taken under polarized light.}\label{HPM}
\end{figure}

The distribution of the $c$-axis component of the DC magnetic flux density at the top (a) and the bottom (b) surfaces of the sample is represented in Figure~\ref{HPM}-(a,b). The dark areas correspond to low magnetic flux densities while the bright areas are related to a high magnetic flux. Although we have measured the distribution of the magnetic flux density at the whole top and bottom surfaces, we only show it above the quarter of the sample where we have performed the measurements of the magnetic flux inside the holes.

First it can be observed that the Hall probe mapping does not resolve spatially the position of the holes. Their position is superimposed on the mappings for clarity. Second, noticeable differences appear between the bottom and the top surfaces. Not only the values of the magnetic flux density differ in magnitude (as commonly observed with top-seeded melt-grown YBCO, see e.g. Ref~\cite{F2}), but also the distribution of the magnetic flux density varies between the top and bottom surfaces. In particular, one can distinguish, at the top surface, a $\supset$-shaped shielded area with low magnetic flux density, whereas the low-field area at the bottom surface exhibits a more regular (circular) shape. 

The distribution of the magnetic field at both surfaces is different from what would be expected in a sample without defects~\cite{19}. The peculiar distribution observed at both surfaces most likely arises from cracks, defects or disoriented grains~\cite{H1,G7}. In order to further characterize the sample, we have performed optical polarized light micrographs that are reproduced in Figure~\ref{HPM}-(c,d) for the quarter of the top (c) and bottom (d) surfaces. These micrographs do not underline the presence of multiple superconducting grains with different orientations, but indicate the presence of cracks on the bottom surface (especially at the top right corner and at the bottom of the picture). However, the micrographs do not clearly point to either cracks or multiple superconducting grains that could explain the observed distribution of the magnetic flux density. Moreover, it is impossible to know from such surface characterization techniques (Hall probe mapping and microscopy) whether these inhomogeneities extend or not inside the bulk of the sample. The signals from coils placed inside the holes, as reported in the next section, can then be used to gain further information on the presence of invisible inhomogeneities in the sample.

\section{Experimental method for measuring the magnetic field inside the holes}
\label{s:exp}

We use micro coils to measure the magnetic field threading the holes of the drilled melt-textured sample. The sample was immersed in liquid nitrogen ($T = 77~\mathrm{K}$) and the cooling procedure was always performed in zero field (Zero Field Cooled - ZFC). Once cooled, the sample was subjected to a uniform AC magnetic field, applied parallel to its c-axis (i.e. parallel to the axis of the holes). The inductive electromotive force (emf) appearing across each micro coil was recorded to determine the magnetic flux density threading the corresponding hole. 

In order to record a measurable electromotive force in most holes, the experiment has to be carried out with AC magnetic fields of sufficiently large amplitudes, reaching the penetration field, $\mu_0H_p$, of the sample. In the present case, the full penetration field $\mu_0H_p$ is of the order of $100~\mathrm{mT}$; such a large value requires a specifically designed experimental set-up. Several home-made systems use an air coil for generating AC fields with an amplitude up to $100~\mathrm{mT}$~\cite{E2,I1,I2,I3,I4}. In our experiment, we use an iron-core electromagnet in order to generate magnetic fields with a larger amplitude than $100~\mathrm{mT}$. Three 1000-turns coils are wound around a 'C-shape' laminated core, consisting of a stack of Si-doped thin iron sheets ($0.3~\mathrm{mm}$ thickness) which are isolated from one another in order to reduce eddy currents~\cite{I5}. The core has a cross-section of $7~\mathrm{cm}\times8~\mathrm{cm}$. The sample holder is inserted in a pyrex cryogenic vessel placed in the air gap of the electromagnet (width of $2.4~\mathrm{cm}$). The large size of the core cross-section guarantees an excellent homogeneity of the applied magnetic field over the sample volume. In our system, the field inhomogeneity over the volume of the sample was measured to be less than $1~\%$. 

\begin{figure}[t]
\center
\includegraphics[width=6cm]{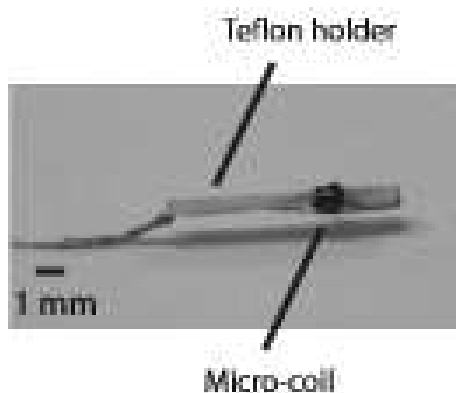}\caption{One of the micro-coil inserted in the holes of the drilled YBCO sample. The coil consists of 10 turns of insulated copper wire with a diameter of $50~\mu\mathrm{m}$. The wire is wound on an insulating cylinder whose diameter is $0.5~\mathrm{mm}$.}\label{coil}
\end{figure}

The micro coils are made of a single layer of 10 turns of fine insulated copper wire ($\phi=50~\mu\mathrm{m}$) tightly wound on an insulating Teflon cylinder with a diameter of $0.5~\mathrm{mm}$ (Figure~\ref{coil}). The micro coils are either positioned symmetrically with respect to the median plane of the sample (centre position), or placed at less than $1~\mathrm{mm}$ from the top or bottom surfaces of the sample. Since the emf appearing across each coil probes the magnetic flux averaged over the coil volume, we use a number of turns (10) that is small enough to ensure a good spatial resolution, while yielding a measurable signal. 

The wires connecting the micro coils to the measuring devices are finely twisted together in order to reduce a spurious inductive pick-up. At $B = 100~\mathrm{mT-RMS}$ (except when explicitly stated, the AC quantities are expressed in RMS values), $f = 25~\mathrm{Hz}$ and $T = 300~\mathrm{K}$, the emf produced by the magnetic flux through the twisted wires is measured to be $\sim1~\mu \mathrm{V}$. This value is around 5\% of the emf developed across a 10-turn coil under the same conditions. The emf of the micro coils is measured by both (i) a SR560 low-noise preamplifier followed by a TDS 2012B digitizing oscilloscope in order to record its waveform and (ii) a Perkin Elmer 7260 lock-in amplifier in order to characterize its fundamental harmonic. The lock-in amplifier requires a stable reference signal. In commercial systems as well as those reported in the literature~\cite{D1,I6,I7}, it is a common practice to take as a reference the voltage drop across a precision resistor placed in series with the magnet coils. In the present case, however, this procedure cannot be used because the low-value resistor ($< 1~\Omega$) required to sustain the high AC currents would exhibit a small (but finite) reactance, giving rise to a parasitic phase-lag. In our experiment, the reference signal is the emf across an 50-turn coil wound around the iron core; this emf is phase shifted by exactly $90^\circ$ with the applied AC magnetic field. 

Prior to every measurement sequence, a calibration procedure of the micro coils was performed at room temperature: the micro coils were positioned in the holes, the sample placed in the cryogenic vessel inside the air gap of the electromagnet, and subjected to a known magnetic field (RMS value $B$, frequency $f$). The RMS value of the emf of the coil ($N$ turns), $V = 2\pi N f B A$, was measured by the lock-in amplifier to determine the effective cross-section A of each micro coil. Note that the effective cross-section is found to match the geometric cross-section of the teflon holder within $80~\%$. The difference can be attributed to the finite diameter of the wire and to a slight tilting of the winding with respect to the applied field. That effective cross-section is then used to convert the measured emf into a magnetic flux density.

\section{Results and discussion}
\label{s:result}

\subsection{Time evolution of the emf across the micro coils}

\begin{figure}[p]
\center
\includegraphics[width=10cm]{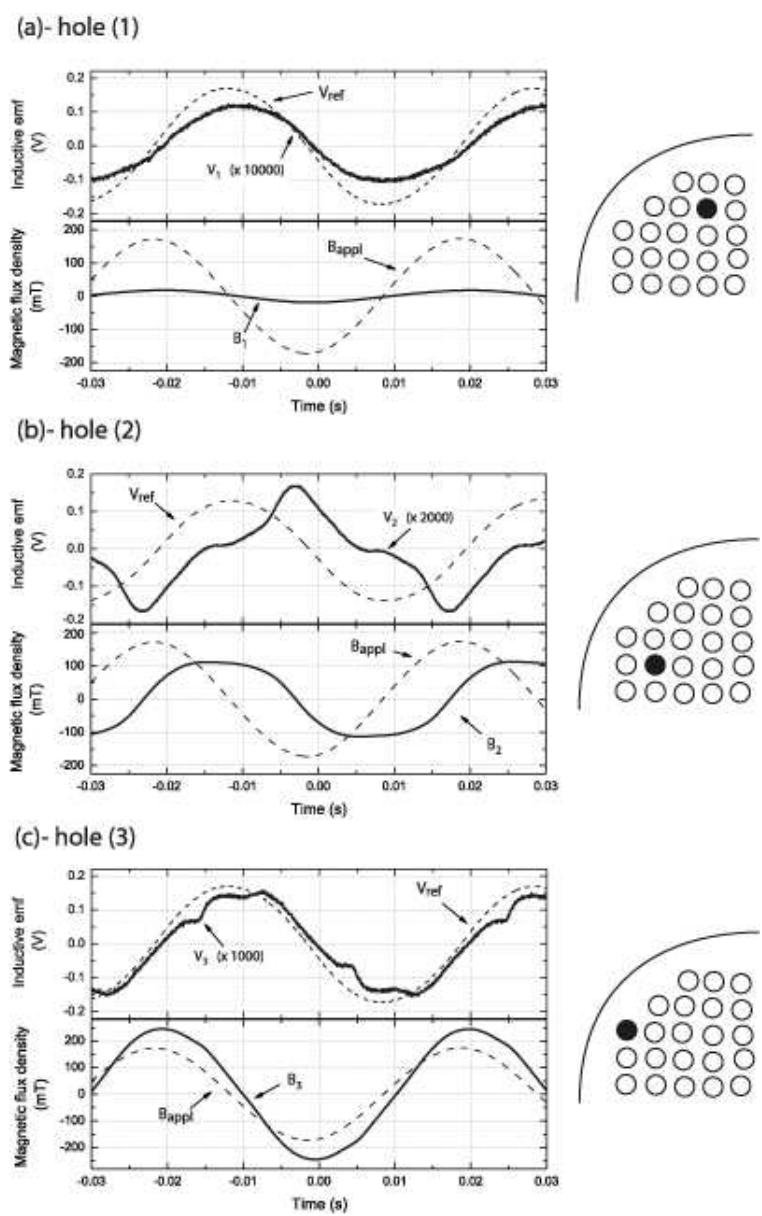}\caption{(i) Top halves of the graphs: Inductive electromotive force ($V_1$, $V_2$ and $V_3$ - solid line) vs. time as measured with a micro coil inserted (in the central position) inside hole (1) , (2) and (3), respectively (see Figure~\ref{HPM}-(a)). The reference emf ($V_{\mbox{\footnotesize{ref}}}$ - dashed line) is at $90^\circ$ with respect to the sinusoidal applied field ($120~\mathrm{mT}$ at $24.73~\mathrm{Hz}$). (ii) Bottom halves of the graphs: Magnetic flux density vs. time as obtained by a numerical integration of the inductive emf. Dashed lines: applied magnetic flux density, $B_{\mbox{\footnotesize{appl}}}$. Solid lines: magnetic flux density, $B_1$, $B_2$ and $B_3$, measured inside holes (1), (2) and (3). }\label{signals}
\end{figure}

Let us first consider the time evolution of the inductive emf, together with the resulting magnetic flux densities, as measured by three micro coils inserted inside hole (1), (2), and (3), respectively. These holes are located in three particular regions determined from the magnetic flux distribution at the top surface shown in Figure~\ref{HPM}-(a). Hole (1), (2) and (3) are respectively located in a low, intermediate, and high magnetic flux region. The micro coils are centered in their corresponding hole. The applied magnetic flux density is $120~\mathrm{mT}$ and has a frequency of $24.73~\mathrm{Hz}$. The upper halves of Figures~\ref{signals} (a), (b) and (c) show the time evolution of the inductive emf ($V_1$, $V_2$, $V_3$ - solid line) in holes (1), (2) and (3). For the sake of clarity, the inductive emf's are shown after amplification by the low-noise preamplifier, the gain being indicated in brackets. The reference emf ($V_{\mbox{\footnotesize{ref}}}$ - dashed lines) corresponds to the inductive emf across a 50-turn coil wound around the iron core. 

Three different waveforms can be distinguished in the hole (1), (2), and (3), respectively.
\begin{enumerate}
\item In hole (1), the inductive emf across the micro coil is a sine wave. There is a negligible phase-shift of $V_1$ with respect to the reference signal $V_{\mbox{\footnotesize{ref}}}$. The emf in this case is to be attributed to a parasitic emf produced by the applied field that crosses closed loops of wires connecting the coil. 
\item In hole (2), the inductive emf is strongly distorted, and exhibits a succession of positive and negative peaks, reminiscent of those observed with a pick-up coil inserted inside a HTS tube made of polycrystalline Bi-2223 subjected to an axial AC magnetic field~\cite{tubes}.
\item The inductive emf produced by the micro coil inserted inside hole (3) exhibits a slightly distorted sine shape with almost no phase-shift with respect to the reference emf. 
\end{enumerate}

Let us now focus on the magnetic flux density inside the hole. The flux densities are determined by numerical integration of the inductive emf's shown in the top panels of Figure~\ref{signals}. The bottom panels of Figures~\ref{signals} show the time evolution of the magnetic flux density ($B_1$, $B_2$, $B_3$ - black line) crossing the median section of the hole. The applied magnetic flux density ($B_{\mbox\footnotesize{appl}}$ - dashed line) is also shown for comparison.

The time dependences of the magnetic flux density are different in each of the three holes and are correlated to the respective emf waveforms measured inside these holes. 
\begin{enumerate}
\item In hole (1), the magnetic flux density is a sine wave whose amplitude is much smaller than the applied one (by a factor of~$\approx10$).
\item The magnetic flux density corresponding to the strongly distorted emf in hole (2) is a clipped sine wave that lags the applied field in time. Such characteristics are the consequence of strong pinning in the sample which leads to a hysteresis between the applied field and the local induction in the hole. That behaviour is qualitatively consistent with the Bean model~\cite{19}.
\item In hole (3), the magnetic flux density is almost sinusoidal and has an amplitude larger than the applied magnetic flux density (by a factor of~$\approx1.5$).
\end{enumerate}

It is emphasized that the signals measured in the other holes, although not discussed in this work, are similar to those presented in Figure~\ref{signals}.

\subsection{Classification of three different penetration regimes}

We now analyse the magnetic flux density inside the hole (1), (2) and (3) (as indicated in Figure~\ref{signals}) as a function of the applied magnetic flux density. The applied magnetic flux density varies from $30$ to $130~\mathrm{mT}$. The micro coil was placed as carefully as possible in a central position in each case.

Figure~\ref{3behav} shows the amplitude of the fundamental harmonic of the magnetic flux density inside the holes (symbols + solid lines), as a function of the applied magnetic flux density (dashed lines). Quite similarly to the waveform pattern, we can distinguish three different behaviours (that are further referred to as ''penetration regimes''), as a function of the hole location in the sample:

\begin{figure}[t]
\center
\includegraphics[width=12cm]{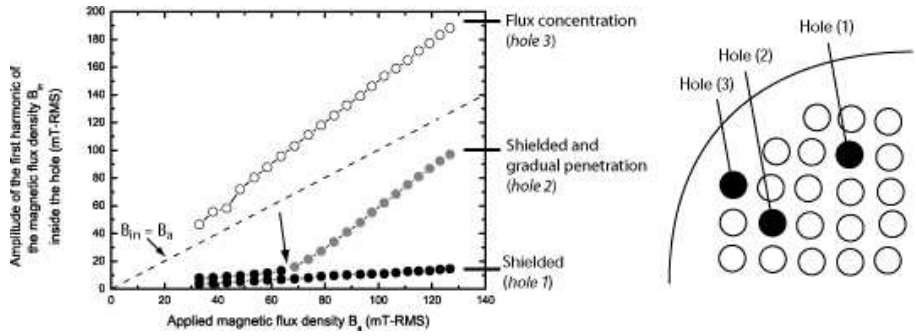}\caption{Amplitude of the fundamental harmonic of the magnetic flux density inside hole (1), (2), and (3) (their position in the sample is reminded in the right panel) as a function of the applied magnetic flux density varying from $30$ to $130~\mathrm{mT}$. The arrow shows the transition between the shielded regime and the gradual penetration regime.}\label{3behav}
\end{figure}

\begin{enumerate}
\item {\it the shielded regime}, for which the magnetic flux inside the hole is negligible with respect to the applied one (attenuated by a factor of~$\approx10$). The magnetic flux increases almost linearly with the applied field. Hole (1) is in this regime over the range of applied fields, while hole (2) is in the shielded regime for the lower fields (black circles). The waveform of the emf in this regime is similar to what is observed in Figure~\ref{signals}-(a).
\item {\it the gradual penetration regime}: the hole is penetrated by the magnetic flux, with a density that is smaller than the applied one and that is not so strongly attenuated than in the shielded regime. The magnetic flux density in the hole increases almost linearly with the applied one but with a large slope ($~\approx1.5~\mathrm{mT/mT}$; the slope is $\approx 10$ times larger than in the shielded regime). The observable kink (see arrow in Figure~\ref{3behav}) due to the change of slope in hole (2) defines the transition from the shielded regime to the gradual penetration one (gray circles). 
\item{\it the flux concentration regime}: the hole is also penetrated by the magnetic flux, but now with a density that exceeds significantly the applied one ($\approx 1.5$ times larger). There is thus a flux concentration in the hole, as shown in hole (3) (white circles). 
\end{enumerate}

In the other holes, the behaviour of the fundamental component of the magnetic flux density as a function of the applied field is always consistent with the regimes described above.

\subsection{Distribution of the penetration regime of the holes}

Based on the observations presented in the above paragraphs, we now determine the penetration regime of the holes located in the quarter of sample under consideration at a given applied field. A relevant question is to determine how the distribution of the penetration regimes is modified when the coil is placed in the median plane of the sample and close to the top and bottom surfaces. 

\begin{figure}[b]
\center
\includegraphics[width=12cm]{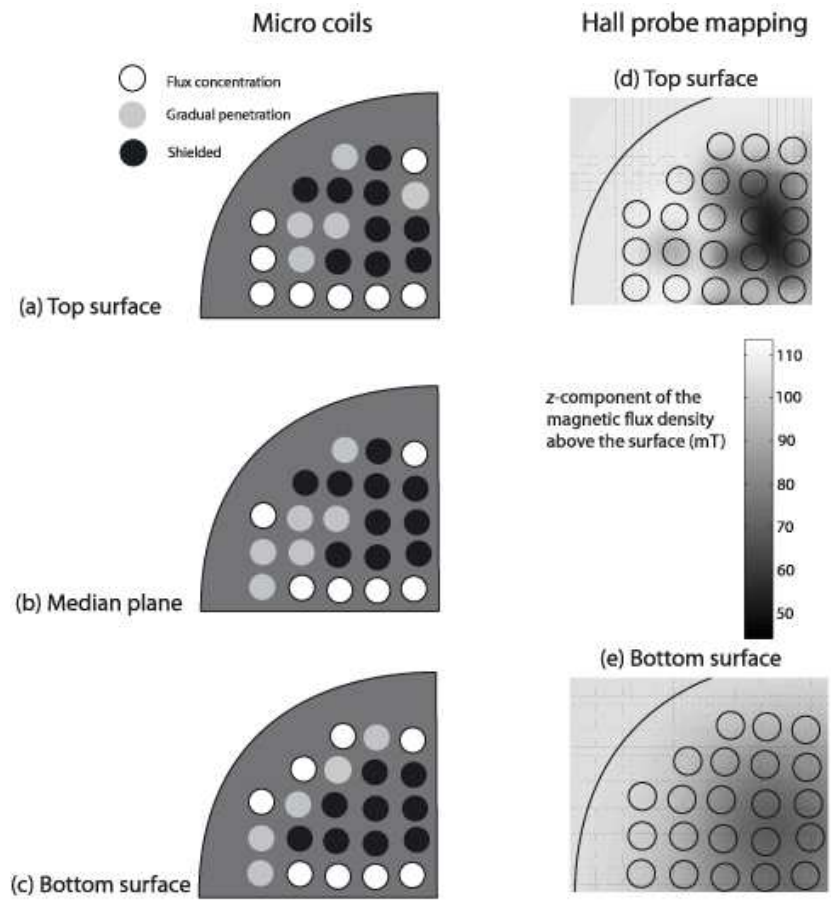}\caption{(i)- Left panels: distribution of the penetration regimes inside the holes of the quarter of the sample for an applied magnetic field of $\mu_0H=100~\mathrm{mT}$ and for micro coils located close to the top surface (a), symmetrically around the median plane (b), and close to the bottom surface (c). (ii)- Right panels: distribution of the magnetic flux density at the sample surfaces (d,e) obtained with the Hall probe mapping discussed in Figure~\ref{HPM}-(a,b).}\label{BehavSummary}
\end{figure}

The distribution of the penetration regimes is shown in Figure~\ref{BehavSummary} for an applied magnetic flux density of $100~\mathrm{mT}$. The micro coils were located close to the top surface in Figure~\ref{BehavSummary}-(a), close to the bottom position in Figure~\ref{BehavSummary}-(b), and in the centre position in Figure~\ref{BehavSummary}-(c). The black holes are in the shielded regime, the gray holes in the gradual penetration regime, and the white holes in the high flux regime.

Independent of the vertical position of the coil inside the hole, it is observed that the magnetic field does not penetrate in the same way the holes that are located at a same radial distance from the border. The holes that are shielded at their top surface or in their centre form a $\supset$-shape which is localised at the right corner of the picture. In a homogeneous sample, one would expect the shielded holes to be in a circular region having the same center as the sample cross-section, whose radius decreases as the applied field increases, until the sample is fully penetrated. The holes outside that shielded region would be in the gradual penetration regime. In a non-homogeneous sample, as studied here, the situation is different.

The distribution of the penetration regimes of the holes is correlated with the distribution of the magnetic flux density at the sample surfaces, shown on the right panels of Figure~\ref{BehavSummary}. As a reminder, the Hall probe mapping is performed under an applied $100~\mathrm{mT}$ DC field. In view of a fair comparison, the distributions of the penetration regimes of the holes are measured under an AC same magnetic field equal to $100~\mathrm{mT-RMS}$. 

A comparison of AC and DC measurements brings out three pieces of information. First, from the Hall probe mapping, we observe that the holes in the shielded regime lie at the surface in a low magnetic flux density region. In particular the $\supset$-shape of low magnetic field at the top surface corresponds in the AC measurements to a region of shielded holes. Second, the holes in the gradual penetration regime (AC experiment) are located in regions of intermediate magnetic flux densities in the surface (DC experiment). Third, it can be observed that the smooth distribution of the penetration regime of the holes close to the bottom surface (AC experiment) is consistent with the distribution of the magnetic flux at that surface (DC measurement).

This comparison illustrates the interest of probing the field in the bulk. The Hall probe mapping technique by itself does not bring enough information to infer the penetration pattern in the volume of the sample. Only the micro coil technique allows one to understand how the magnetic flux penetrates the bulk of the sample. For instance, Figure~\ref{BehavSummary} shows that the field distribution in the median plane is more similar to the one observed close to the top surface than close to the bottom surface. These results are related to the detailed microstructure of the bulk of the sample.

Note that, the flux concentration regime is not observed in either of the Hall probe mappings. In these experiments, the probe, which measures the vertical component of the magnetic field, is located at a fixed distance from the surface, in the spreading out region of the magnetic field lines that are concentrated inside the hole. The flux concentration can only be observed when the flux is probed inside the holes. Interestingly, the holes with a flux concentration (especially close to the top surface) surround more efficiently shielded regions. These observations suggest that the flux concentration results from demagnetization effects associated to the microstructure of the sample. Cracks or defects present obstacles for the current flow in the sample. These obstacles cause an irregular distribution of the magnetic flux, with an apparition of shielded regions that are not necessarily located in the centre of the sample. The return path for the demagnetizing field coming from a given shielded region crosses the neighbouring holes and the demagnetizing field increases locally the magnetic flux density inside these holes. Such demagnetizing effects are also responsible for the flux concentration inside natural or artifical grain boundaries~\cite{H1,anisotropy}.

\section{Conclusions}

The micro coil measurement technique is appropriate for a local characterization of the magnetic properties of drilled samples. 

We measured the magnetic flux density inside the bulk of a drilled YBCO cylinder with the help of micro coils inserted in the holes. The sample was subjected to a uniform AC magnetic field ranging from $30$ to $130~\mathrm{mT}$. The inductive electromotive force across the micro coil was recorded with an oscilloscope and the amplitude of its fundamental harmonic was measured using a lock-in amplifier. Three penetration regimes are observed: (i)-the shielded regime, for which the magnetic flux does not penetrate the hole and its time evolution exhibits a sine shape of very small amplitude; (ii)- the gradual penetration regime, for which the magnetic flux density in the hole exhibits a clipped sine shape and the amplitude of its fundamental harmonic increases linearly with the applied field, the pick-up signal in this case contains a succession of peaks; (iii)- the flux concentration regime, for which the magnetic flux density in the hole is larger than the applied one and the pick-up voltage is almost a sine-wave. 

The distribution of the penetration regimes is closely related to the magnetic flux density measured at the surfaces by a DC Hall probe mapping. Knowledge of this distribution for various vertical positions of the coil gives information on the bulk magnetic properties of the sample and of its microstructure. The local probing of the magnetic flux density inside the holes brings pieces of information that are impossible to obtain with usual surface characterization techniques. The analysis of the signals from the coils placed inside the hole allows one to know how a flux distribution at the surface is modified in the bulk and to measure flux concentration appearing in the volume of an inhomogeneous sample.

\section{Acknowledgments}

We thank the \textit{Fonds de la Recherche Scientifique (FRS-FNRS)} of Belgium, the University of Li\`ege, and the Royal Military Academy of Belgium for cryofluid and equipment grants.

\end{document}